\documentclass[11pt,a4paper]{article}
\usepackage{jheppub}
\usepackage{amssymb} \usepackage{commath}
\usepackage{amsmath}
\usepackage{url}
\usepackage{float}
 \usepackage[utf8]{inputenc}
 \usepackage{tikz}
\usepackage[T1]{fontenc} \usepackage{lmodern}
\usepackage{braket}
\usepackage{graphicx}
\usepackage{pictexwd,dcpic}
\usepackage{braket}
\usepackage{atbegshi,picture, afterpage}
\usepackage{lipsum}
\usepackage{physics}
\usepackage{appendix}
\usepackage{bbm}
\usepackage{hyperref}
\usepackage{color}
\usepackage{bm}
\usepackage[makeroom]{cancel}
\usetikzlibrary { decorations.pathmorphing, decorations.pathreplacing, decorations.shapes,
}

\newcommand{\be}{\begin{equation}}
\newcommand{\ee}{\end{equation}}
\newcommand{\ba}{\begin{align}}
\newcommand{\ea}{\end{align}}
\usepackage{amssymb} \usepackage{commath}
\begin{document}
\title{Hamiltonian charges in the asymptotically de Sitter spacetimes}
\author{Maciej Kolanowski}
\author{and Jerzy Lewandowski}
\affiliation{Institute of Theoretical Physics, Faculty of
  Physics, University of Warsaw, Pasteura 5, 02-093 Warsaw, Poland}
  \emailAdd{Maciej.Kolanowski@fuw.edu.pl}
  \emailAdd{Jerzy.Lewandowski@fuw.edu.pl}
  \date{\today}

\abstract{
We generalize a notion of 'conserved' charges given by Wald and Zoupas to the asymptotically de Sitter spacetimes. Surprisingly, our construction is less ambiguous than the one encountered in the asymptotically flat context.  An expansion around exact solutions possessing Killing vectors provides their physical meaning. In particular, we discuss a question of how to define energy and angular momenta of gravitational waves propagating on Kottler and Carter backgrounds. We show that obtained expressions have a correct limit as $\Lambda \to 0$. We also comment on the relation between this approach and the one based on the canonical phase space of initial data at $\mathcal{I}^+$.}
\maketitle
\section{Introduction} \noindent 
Energy and momenta of gravitational waves in the full non-linear general relativity were first obtained by Trautman over 60 years ago \cite{Trautman:2016xic} with his work followed by nowadays standard approach of Bondi et al. \cite{Bondi:1962px}\footnote{For a more detailed history, see \cite{DensonHill:2016upp}.}. These results were further refined and extended to the angular momenta in the eighties \cite{Penrose:1982wp,Dray:1984rfa, Shaw:1984sfa}. \\
Independently, Hamiltonian description of GR was investigated \cite{Ashtekar79, Kijowski:1979dj}. It was applied to the problem of radiation in \cite{Ashtekar:1981bq, Chrusciel:2002cp}. Symplectic approach is endowed with certain ambiguities -- one is allowed to change Hamiltonian by a boundary term if it is followed by an analogous change in the flux formula. Thus, an additional input is needed to fix physical values of local fluxes. (Since the change is only in boundary terms, the global quantities are unchanged.) Authors of \cite{Ashtekar:1981bq} carefully chose  topology on the phase space to recover Trautman-Bondi mass loss formula. Another solution to this problem was offered in \cite{Wald:1999wa} (which followed \cite{Iyer:1994ys}) where an algorithm for fixing boundary terms was proposed in terms of the presymplectic potential $\Theta$ at $\mathcal{I}^+$ (or a boundary in general). In this way, the ambiguity of boundary terms was reformulated as an ambiguity in the choice of $\Theta$. It was shown that upon certain additional restrictions, it is a unique object and thus this algorithm gives an unambiguous answer which coincides with the results of \cite{Ashtekar:1981bq}. Perhaps a more natural version of this approach was recently presented in \cite{Harlow:2019yfa}. \\
All those important developments started with an assumption that a cosmological constant $\Lambda$ is vanishing. However, observations show $\Lambda > 0$ \cite{riess1998observational}. The question of how to define the energy of gravitational waves (GW) in the presence of such $\Lambda$ attracted a lot of attention in the last few years (see e.g. \cite{kastor2002positive, Penrose:2011zza, Penrose:2011zza, Ashtekar:2015lla, Szabados:2015wqa, Chrusciel:2020rlz, Ashtekar:2014zfa, He:2015wfa, Poole:2018koa, Compere:2019bua, Ashtekar:2015lxa, Bishop:2015kay, Date:2015kma, Kolanowski:2020wfg} for different definitions, proposal, arguments and calculations.) Since null infinity $\mathcal{I}$ is spacelike in asymptotically de Sitter spacetimes, this task is much harder -- no natural timelike asymptotic symmetry is available. Moreover, there is virtually no universal structure at $\mathcal{I}$ (beyond that of a smooth manifold) which could be used. In this work, we generalize Wald-Zoupas notion of 'conserved' charges to the case $\Lambda > 0$. Surprisingly, one finds that the answer to be much less ambiguous than in the asymptotically flat spacetime.    \\
It is highly non-obvious whether those charges could be interpreted as energy, momenta and angular momenta because there is no natural way to distinguish vector fields that could generate those particular physical quantities. Nevertheless, we show on examples that expansion of our results around an exact solution equipped with a Killing vector has a natural interpretation of a charge associated with that vector field.\\ Those formulas for charges and fluxes were obtained before in a more abstract context \cite{Anninos:2010zf}. However, to our best knowledge, this note contains the first proof that these results are unambiguous. Similar work was also done in \cite{Balakrishnan:2019zxm} but with much more severe boundary conditions, which excluded gravitational waves carrying any de Sitter charges. There is another Hamiltonian approach, based on the initial data at $\mathcal{I}^+$, given by Friedrich \cite{friedrich1986}, with a suitable fall-off as one approaches $i^o$ and $i^+$ along $\mathcal{I}^+$ \cite{aa-sanya}. Thanks to these boundary conditions, it is possible to distinguish between gauge and symmetry vector fields on $\mathcal{I}^+$ and associate \emph{global} fluxes associated with symmetries across all of $\mathcal{I}^+$. It was pointed out in \cite{aa:ropp} that the expression of \emph{local} fluxes associated with arbitrary vector fields on $\mathcal{I}^+$ have certain physically undesirable features. The global fluxes associated with symmetries are free from this drawback. Relation between the framework presented in this paper and that summarized in \cite{aa-sanya} will be discussed in a forthcoming publication.
 \\
The rest of the paper is organized as follows.
In Sec. \ref{dS} we introduce a necessary theory of the asymptotically de Sitter spacetimes. In Sec. \ref{wz} we quickly summarize the Wald-Zoupas notion of 'conserved' charges. We stick to the conventions of \cite{Wald:1999wa} to avoid confusion. Then, in Sec. \ref{gen}, we generalize Wald-Zoupas charges to this context and prove their uniqueness. We discuss their physical meaning in Subsection \ref{Special}. The main results are repeated and discussed in Sec. \ref{con}. Technical details are relegated to the Appendix.

\section{Asymptotically de Sitter spacetimes} \label{dS}
 \noindent
We will start with a quick review of the asymptotically de Sitter spacetimes. This is going to be rather a practical introduction to the topic, for a more thorough discussion see \cite{Ashtekar:2014zfa}. For simplicity we restrict ourselves to the vacuum Einstein equations in four dimensions:
\begin{equation}
    R_{\mu\nu} = \Lambda g_{\mu\nu} \label{ee}
\end{equation}
with $\Lambda = \frac{3}{\ell^2} > 0$. Inclusion of the matter fields would not be very hard but one should be a little bit careful with assumed asymptotics of a stress-energy-momentum tensor which could alter the final results. An asymptotically de Sitter spacetime near\footnote{the same is true for $\mathcal{I}^-$, we choose $\mathcal{I}^+$ only for convenience} $\mathcal{I}^+$ can be put into a Fefferman-Graham gauge\cite{Starobinsky:1982mr,AST_1985__S131__95_0}:
\begin{equation}
    g = - \frac{\ell^2 d\rho^2}{\rho^2} + \frac{\ell^2 dx^a dx^b}{\rho^2} \sum_{i=0} g^{(i)}_{ab} \rho^i, \label{FG}
\end{equation}
where $\rho = 0$ corresponds to the null infinity $\mathcal{I}^+$ which is spacelike. Tensors $g_{ab}^{(i)}$ are all defined on $\mathcal{I}^+$. Since our considerations are all local, we do not assume anything about the topology of the null infinity. However, one should keep in mind that an algebra of asymptotic symmetries is going to be a little bit different, depending upon that topology. Moreover, most results (like \cite{friedrich1986}) about existence and stability of solutions assume that $\mathcal{I}^+ \sim \mathbb{S}^3$.  \\
From the Einstein equations we obtain:
\begin{align}
    g^{(1)}_{ab} &= 0 \\
    g^{(2)}_{ab} &= \mathring{R}_{ab} - \frac{1}{4}\mathring{R} g_{ab}^{(0)} \\
    g^{(0)ab}g^{(3)}_{ab} &= 0 \label{cons1} \\
    D^a g^{(3)}_{ab} &=0, \label{cons2}
\end{align}
where $\mathring{R}_{ab}$ and $\mathring{R}$ are Ricci tensor and scalar (respectively) of $g^{(0)}$ and $D$ is its covariant derivative. For convenience, let us introduce a holographic stress-energy tensor:
\begin{equation}
    T_{ab} = \frac{\sqrt{3\Lambda}}{16\pi G}g^{(3)}_{ab}
\end{equation}
which can be specified freely up to the constraints $($\ref{cons1}-\ref{cons2}$)$. Expansion \eqref{FG} is not unique, $g_{ab}^{(0)}$ and $T_{ab}$ are defined up to the conformal transformations:
\begin{align}
    \begin{split}
        g^{(0)}_{ab} &\mapsto \Omega^2 g^{(0)}_{ab} \\
        T_{ab} &\mapsto \Omega^{-1} T_{ab},
    \end{split}
\end{align}
which correspond to the choice of $\rho = \textrm{const.}$ slices. $T_{ab}$ can be also easily expressed through the electric part of a Weyl tensor:
\begin{equation} 
    T_{ab} = -\frac{\ell}{8\pi G}\mathcal{E}_{ab} := -\frac{\ell}{8\pi G} \lim_{\rho \to 0} \rho \ell^{-1} C_{a\rho b\rho}, \label{weyl}
\end{equation}
where $C_{\mu \nu \sigma \delta}$ is the Weyl tensor of $g$. It follows from the pioneering work of Friedrich \cite{friedrich1986} that a conformal class $[(g^{(0)}_{ab}, T_{ab})]$ is the full Cauchy data for $g_{\mu \nu}$, at least in the neighborhood of the de Sitter solution. In the case of the gauge \eqref{FG}, it means that $g^{(n)}_{ab}$ ($n>3$) are determined by  $(g^{(0)}_{ab}, T_{ab})$ recursively, no additional integration constants arise. \\
Equations of motion \eqref{ee} follow from the finite action (see \cite{Balasubramanian:1999re, Compere:2019bua} for a derivation using holographic renormalization)
\begin{equation}
    S_{EH} = \frac{1}{16\pi G}\int_M d^4 x \sqrt{|g|} \left( R - 2\Lambda \right) + \frac{1}{16\pi G}\int_{\partial M} d^3 x \sqrt{\gamma} \left(
    2K + \frac{4}{\ell} - \ell R[\gamma]
    \right),
\end{equation}
where $\gamma_{ab}$ is an induced metric on a (spacelike) boundary of $M$ and $K$ is a trace of the second fundamental form of $\partial M$. Counterterms are chosen carefully to render the action finite. Incidentally, it follows from Sec. \ref{uniqueness} below that no additional counterterms are allowed if $\partial M = \mathcal{I}^+$. Then, presymplectic current is finite on the slices $\rho = \textrm{const.}$ Taking the limit $\rho \to 0$ we find
\begin{equation}
    \overline{\boldsymbol{\omega}}(g^{(0)},T;\delta_1,\delta_2) = \frac{1}{\ell^2} \delta_{[1}\left(\sqrt{g^{(0)}} T^{ab} \right) \delta_{2]} g_{ab}^{(0)} d^3 x.
\end{equation}
One can show that integral of $\overline{\omega}$ is equal to the presymplectic form on $\mathcal{F}$.
\section{Wald-Zoupas 'conserved' charges} \label{wz} \noindent 
In this section, we will discuss the general procedure of defining charges using Wald-Zoupas prescription and show it on an example of the vacuum GR at null infinity. For the convenience of the Reader, we will try to follow \cite{Wald:1999wa} as closely as possible. \\
We consider a diffeomorphic invariant theory with fields $\phi$ (such as a metric but possibly also other tensor fields) defined on an $n$ dimensional manifold $M$. We assume that a suitable configuration space $\mathcal{F}$ to which $\phi$ belongs was already chosen. Field dynamics is given by a Lagrangian $n$-form $\bf{L}$ which depends on the fields $\phi$ in a diffeomorphism  covariant way\footnote{Bold font is used to denote that our objects are differential forms in $M$ rather than scalars. }. Variation of $\bf{L}$ leads to a presymplectic potential current $\boldsymbol{\theta}$ by
\begin{equation}
    \delta {\bf L} = {\bf E}(\phi) \delta \phi + d\boldsymbol{\theta}(\phi; \delta \phi),
\end{equation}
where $\bf{E}$ are equations of motions (so ${\bf E}(\phi) = 0$ on shell. We will denote a space of solutions by $\overline{\mathcal{F}} \subset \mathcal{F}$), $d$ stands for the exterior derivative in $M$  and $\delta \phi \in T_\phi \mathcal{F}$. Clearly, such $\boldsymbol{\theta}$ is defined up to an addition of an $(n-1)$-form in $M$ which is a covariant function of $\phi$ and is closed.  $\boldsymbol{\theta}$  can be used to define a presymplectic current $\boldsymbol{\omega}$:
\begin{equation}
    \boldsymbol{\omega} (\phi; \delta_1 \phi, \delta_2 \phi) = 2 \delta_{[1}\boldsymbol{\theta} (\phi; \delta_{2]} \phi),
\end{equation}
that satisfies
\begin{equation}
    d \boldsymbol{\omega} (\phi; \delta_1 \phi, \delta_2 \phi) = 0,
\end{equation}
for every pair of vectors tangent to $\overline{\mathcal{F}}$. 
The current $\bf{\omega}$ in turn leads to  functionals on $\mathcal{F}$ by
\begin{equation}
    \Omega_\Sigma (\phi; \delta_1 \phi, \delta_2 \phi) = \int_\Sigma \boldsymbol{\omega} (\phi; \delta_1 \phi, \delta_2 \phi)
\end{equation}
where $\Sigma$ is any $(n-1)$-dimensional hypersurface such that right hand side exists. In particular, if $\phi \in \overline{\mathcal{F}}$, $\delta_1 \phi, \delta_2 \phi \in T_\phi \overline{\mathcal{F}}$, then $\Omega_\Sigma$ yields the same value for any Cauchy surfaces $\Sigma$ (assuming appropriate fall-off conditions). It is then, a presymplectic form on $\mathcal{\overline{F}}$. (The prefix 'pre-' denotes the fact that such $\Omega_\Sigma$ can be degenerate.) \\
Given $\Sigma$ (not necessarily a Cauchy surface) and a vector field $\xi$, a function $H_\xi [\Sigma]: \mathcal{F} \to \mathbb{R}$ is called a Hamiltonian conjugate to $\xi$ if for all $\phi \in \mathcal{\overline{F}}$ and $\delta \phi \in T_\phi \overline{\mathcal{F}}$:\footnote{Note that in \cite{Wald:1999wa}, it was only assumed that $\delta \phi \in T_\phi \mathcal{F}$. If we found $H_\xi$ on $\mathcal{F}$, we could extend it to $\mathcal{F}$ (if needed) in such a way that the right hand side of \eqref{cancel} held so it does not change anything.}
\begin{equation}
    \delta H_\xi[\Sigma] = \Omega_\Sigma (\phi; \delta \phi, \mathcal{L}_\xi \phi). \label{cancel}
\end{equation}
For  general $\xi$, the right hand side is not a full variation and thus there is no $H_\xi$. To emphasize this fact, we will write $\cancel{\delta}H_\xi$ instead of $\delta H_\xi$. Let us emphasize here that $\cancel{\delta}$ is not an operator, it has no meaning outside of $\cancel{\delta}H_\xi$ which is defined by the right hand side of the Eq. \eqref{cancel}.  In diffeomorphic covariant theories (which we consider here) it can be rewritten as \cite{Iyer:1994ys}
\begin{equation}
    \cancel{\delta} H_\xi [\Sigma] = \int_\Sigma \boldsymbol{\omega} (\phi; \delta \phi, \mathcal{L}_\xi \phi) = \int_{\partial \Sigma} [\delta \bf{Q} - \xi \cdot \boldsymbol{\theta}], \label{boundary}
\end{equation}
where $\bf{Q}$ is Noether charge $(n-2)$ form and $\partial{\Sigma}$ is a boundary of $\Sigma$. We are mainly interested in surfaces $\Sigma$ which stretch up to the null infinity, hence an appropriate limit procedure is needed to define the integral along $\partial \Sigma$. It may happen $\delta \bf{Q}$ and $\boldsymbol{\theta}$ do not possess a limit but this particular linear combination does. An example of such hypersurfaces $\Sigma$ in the asymptotically de Sitter spacetimes is shown in the Fig. \ref{fig:hypersurfaces}. \\
Eq. \eqref{boundary} allows one to easily find an obstruction for $H_\xi$ to exist. Indeed, taking 'the second variation', we obtain
\begin{equation}
        \delta_1 \cancel{\delta}_2 H_{\xi} - \delta_2 \cancel{\delta}_1 H_{\xi}
        = - \int_{\partial\Sigma} \xi \cdot \boldsymbol{\omega}(\phi, \delta_1 \phi, \delta_2 \phi).
\end{equation}
Generically, the right hand side does not vanish and thus $\cancel{\delta}H_\xi$ is not a closed form. Let us also notice that since we do not know anything about topology of $\overline{\mathcal{F}}$, even vanishing of the term involving $\xi \cdot \bf{\omega}$ does not imply that $\cancel{\delta}H_\xi$ is exact -- one needs to check it explicitly.\\
\begin{figure}[t]
    \centering
\begin{tikzpicture}[scale=0.85]
\path
(0,0) + (45:10) coordinate[label=45:$i^0$]  (II)
(0,0) + (135:10) coordinate[label=135:$i^+$]  (I);
\draw[very thick, red] (I) -- node[midway, above, black]{$\mathcal{I}^+$}  (II);
\path (I) -- node[pos=0.3, circle, fill, inner sep =2.5pt, label=below left: $\partial\Sigma_1$](11){} node[pos=0.7, circle, fill, inner sep =2.5pt, label=below right: $\partial\Sigma_1$ ](12){}  (II);
\path (I) -- node[pos=0.1, circle, fill, inner sep=2.5pt, label=below left: $\partial\Sigma_2$](21){} node[pos=0.9, circle, fill, inner sep=2.5pt, label=below right: $\partial\Sigma_2$](22){}  (II);
\path (21) -- node[pos=0.5, above]{$\Delta \mathcal{I}^+$} (11);
\path (12) -- node[pos=0.5, above]{$\Delta \mathcal{I}^+$} (22);
\draw (11) .. controls (0,0) .. (12);
\draw (21) .. controls (0,-1) .. (22);
\path (0,2) -- node[pos=0.0, above]{$\Sigma_1$} node[pos=1, below]{$\Sigma_2$} (0,1);
\end{tikzpicture}
    \caption{An example of hypersurfaces extending to $\mathcal{I}^+$ in an asymptotically de Sitter spacetime. $\Delta \mathcal{I}^+$ is a three dimensional portion of $\mathcal{I}^+$, both $\partial \Sigma_1$ and $\partial \Sigma_2$ are $2$-spheres (each depicted as a pair of bullets). ${\bf F_\xi}$ integrated over $\Delta \mathcal{I}^+$ provides the difference between charges located in $\partial \Sigma_2$ and $\partial \Sigma_1$.}
    \label{fig:hypersurfaces}
\end{figure}
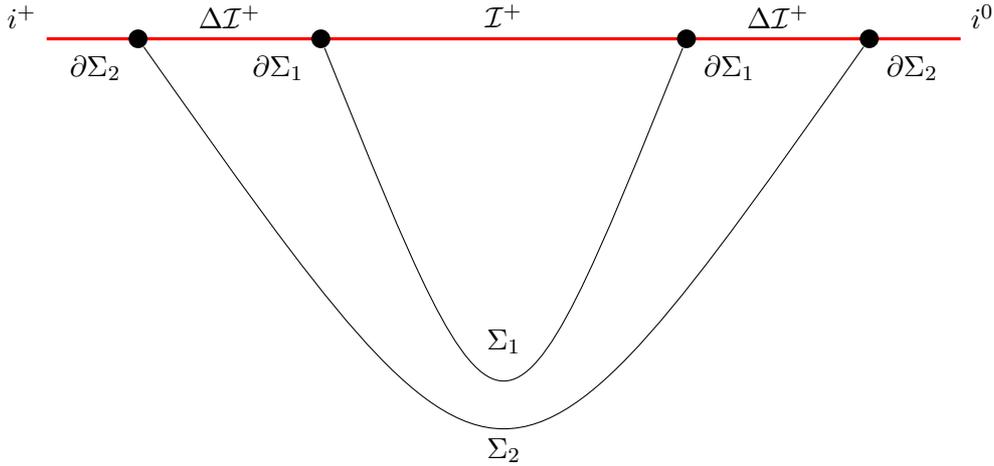Basic idea of \cite{Wald:1999wa} is to add something to the right hand side of Eq. \eqref{boundary} to make it exact. The most obvious choice would be $\boldsymbol{\theta}$. However, as mentioned before, its pullback to $\partial \Sigma$ does not need to exist. Nevertheless, let us assume that pullback of $\boldsymbol{\omega}$ to $\mathcal{I}$ does, let us denote it by $\overline{\boldsymbol{\omega}}$ and by $\bf{\Theta}(\phi; \delta \phi)$ its potential (since $\overline{\boldsymbol{\omega}}$ is defined only on $\mathcal{I}$, so is $\bf{\Theta}$). Then, obviously a variation of
\begin{equation}
    \int_{\partial \Sigma} [\delta {\bf Q} - \xi \cdot \boldsymbol{\theta}] + \int_{\partial \Sigma} \xi \cdot \bf{\Theta}
\end{equation}
vanishes and thus it is a better starting point for the Hamiltonian. In practice, one requires that charges for different choices of $\Sigma$ exist and then they describe e.g. evolving Trautman-Bondi energy which is associated to each cross-section of $\mathcal{I}$. Of course, it is defined only up to a constant which can be fixed on a reference solution $\phi_0$ (e.g. corresponding to the vacuum). The difference of charges for two different cross-sections is flux $\bf{F}_\xi$ through null infinity. An easy calculation shows that
\begin{equation}
    \bf{F}_\xi = \bf{\Theta} (\phi; \mathcal{L}_\xi \phi).
\end{equation}
In general, $\bf{\Theta}$ can be modified by addition of a closed one-form (on $\mathcal{F}$). If we have found such $\bf{\Theta}$ that $\cancel{\delta}H_\xi$ is exact, then this ambiguity is further reduced -- one can only add an exact one-form to $\bf{\Theta}$. The following list of requirements to make it unique was proposed:
\begin{itemize}
    \item[(i)] $\bf{\Theta}$ should be built only out of dynamical fields $\phi$ (or their limits) and an available universal structure at $\mathcal{I}$
    \item[(ii)] $\bf{\Theta}$ should not depend upon any arbitrary choices e.g. a conformal factor
    \item[(iii)] If $\bf{L}$ depends analytically\footnote{One should not mistake an analytical dependence on fields with some restriction on the class of solutions -- those are imposed in the very definition of $\mathcal{F}$} upon $\phi$, then so should $\bf{\Theta}$ 
    \item[(iv)] If $\phi$ is a stationary solution, then ${\bf\Theta}(\phi; \cdot) = 0$.
\end{itemize}
It was found in \cite{Wald:1999wa} that there is only one $\bf{\Theta}$ satisfying all points from the above list and it leads exactly to Trautman-Bondi mass loss formula (and its generalization to the BMS group as calculated in \cite{Dray:1984rfa}). \\
One small element that we somehow glossed over is the choice of a reference solution $\phi_0$ on which all charges are fixed. This choice breaks the covariance of our theory since it distinguished somehow a background field. From the point of view of a full theory, $\phi_0$ is as good as $\psi_\star \phi_0$ where $\psi_\star$ is a diffeomorphism generated by an asymptotic symmetry. Thus, charges should also vanish on  $\psi_\star \phi_0$. This is a non-trivial constraint on $\phi_0$ (or on a theory in general).
\\
In the next section, we will generalize this approach to the asymptotically de Sitter spacetimes. It goes virtually identically, the only change one needs to make regards the list of requirements above. One quickly notices that (iv) is void because there are no stationary spacetimes near $\mathcal{I}^+$ when $\Lambda > 0$. Surprisingly, it will not be needed -- uniqueness follows already from (i-iii). We will also see that the most natural choice of $\phi_0$, namely that all charges vanish on the de Sitter background respects constraints mentioned above. On the other hand, even seemingly stationary solutions like the Schwarzschild-de Sitter spacetime admit non-vanishing fluxes of at least certain asymptotic symmetries. The physical meaning of those is unclear to us at this point.
\section{Generalization to the asymptotically de Sitter spacetimes} \label{gen}
\subsection{Charges} \noindent
We will now proceed according to the Wald-Zoupas prescription. Since the only universal structure in the problem at hand is that of a smooth manifold $\mathcal{I}^+$, asymptotic symmetries are all diffeomorphisms of null infinity. Thus, we want to calculate flux and Hamiltonians associated to all vector fields on $\mathcal{I}$. The (pseudo-)variational expression for the Hamiltonian \eqref{boundary} gives
\begin{equation}
    \cancel{\delta}H_\xi = \int_{\partial\Sigma} d^2 x \left(\delta\left(\sqrt{\sigma} n^a \xi^b T_{ab}\right)
    - \frac{1}{2\ell^2} \sqrt{g^{(0)}} n^a \xi_a T^{bc} \delta g_{bc}
    \right),
\end{equation}
where $\sigma$ is a metric induced on $\partial\Sigma$ and $n^a$ is normal to it. From the form of $\overline{\omega}$ we see that 'the second variation' of $H_\xi$ does not vanish. Thus, we need to add a presymplectic potential ${\bf\Theta}$. The easiest (and, as we will see, the only) choice is:
\begin{equation}
    {\bf\Theta} = \frac{1}{2\ell^2}T^{ab} \delta g^{(0)}_{ab} dVol. \label{theta}
\end{equation}
Thus, we have
\begin{equation}
    \delta H_\xi = \cancel{\delta} H_\xi + \int_{\partial \Sigma} \xi \cdot {\bf \Theta}  = \int_{\partial\Sigma} d^2 x \delta\left(\sqrt{\sigma} n^a \xi^b T_{ab}\right).
\end{equation}
Integration is trivial and it yields
\begin{equation}
H_\xi =     \int_{\partial\Sigma} d^2 x \sqrt{\sigma} n^a \xi^b T_{ab}, \label{hamiltonian}
\end{equation}
where we have chosen integration constants in such a way that all Hamiltonians vanish on the de Sitter solution (on which, in particular, $T_{ab} = 0$). Of course, equation $T_{ab} = 0$ is diffeomorphism-invariant and thus charges vanish on all solutions connected to our choice of the vacuum by an action of an asymptotic symmetry. Flux is given by
\begin{equation}
    {\bf F}_\xi = {\bf\Theta} = \frac{1}{2\ell^2} T^{ab} \mathcal{L}_\xi g^{(0)}_{ab} dVol \label{flux}
\end{equation}
Note that $H_\xi$ and ${\bf F}_\xi$ are conformally invariant, as one would expect. Moreover, for $\int_{\mathcal{I}^+} {\bf F}_\xi$ to be finite, $\xi$ must be a conformal Killing vector field near $i^0$ and $i^+$. \\
Let $\lbrace \cdot, \cdot \rbrace$ be the Poisson bracket obtained from $\overline{\omega}$. It is clear that for an arbitrary slice $H_\xi$ is not a gauge-invariant quantity since gauge transformations correspond to diffeomorphisms which can move the slice. However, if one restricts themselves to the gauge transformations which do not move $i^0$ and $i^+$ (in the spirit of \cite{aa-sanya}), hamiltonians associated with those ends are well-defined observables. In particular their difference, which is equal to the integral of ${\bf F}$, is gauge invariant and it make sense to calculate its Poisson bracket with different observables. Let $\xi, \xi'$ be two vectors fields on $\mathcal{I}^+$ which become conformal Killing vector fields near $i^0$ and $i^+$. An easy calculation shows that on shell
\begin{equation} 
    \left\lbrace \int_{\mathcal{I}^+} {\bf F}_\xi, \int_{\mathcal{I}^+} {\bf F}_{\xi'} \right\rbrace = \int_{\mathcal{I}^+} {\bf F}_{[\xi,\xi']}. \label{pb_flux}
\end{equation}
Thus, no central extension appears in the algebra of fluxes, similarly to \cite{Compere:2020lrt}. Since
\begin{equation}
    \int_{\mathcal{I}^+} {\bf F}_\xi = H_\xi [i^+] -  H_\xi [i^0] \label{diff}
\end{equation}
we can use \eqref{pb_flux} to determine Poisson bracket between charges. Since fields at $i^0$ and $i^+$ are independent, we have
\begin{equation}
    \left\lbrace H_\xi [i^0], H_{\xi'} [i^+] \right\rbrace = 0.
\end{equation}
This implies that
\begin{equation}
     \left\lbrace H_\xi [i^0], H_{\xi'} [i^0] \right\rbrace +  \left\lbrace H_\xi [i^+], H_{\xi'} [i^+] \right\rbrace =  \left\lbrace H_\xi [i^+] - H_\xi [i^0], H_{\xi'} [i^+] - H_{\xi'} [i^0]. \right\rbrace
\end{equation}
Using \eqref{pb_flux} and \eqref{diff} we see that
\begin{equation}
    \left\lbrace H_\xi [i^0], H_{\xi'} [i^0] \right\rbrace +  \left\lbrace H_\xi [i^+], H_{\xi'} [i^+] \right\rbrace = H_{[\xi, \xi']} [i^+] -H_{[\xi, \xi']} [i^0].
\end{equation}
Since we can take in particular $\xi$ and $\xi'$ supported either on $i^0$ and $i^+$, we find that
\begin{align}
    \begin{split}
        \left\lbrace H_\xi [i^0], H_{\xi'} [i^0] \right\rbrace &= -H_{[\xi, \xi']} [i^0] \\
        \left\lbrace H_\xi [i^+], H_{\xi'} [i^+] \right\rbrace &= H_{[\xi, \xi']} [i^+].
    \end{split}
\end{align}
One could be a little bit surprised that although Hamiltonians depend only upon a vector field at $i^0/i^+$, their Poisson bracket involves  their derivatives. However, $\xi$ is asymptotically a conformal Killing vector, its transverse derivatives are also determined.
\subsection{Uniqueness} \label{uniqueness} \noindent
We now want to show that our choice of $\bf{\Theta}$ is unambiguous. We impose the following requirements:  
\begin{itemize}
    \item[(i)] $\bf{\Theta}$ should be built locally (it means in a diffeomorphic covariant manner) only out of $g^{(0)}_{ab}$ and $T_{ab}$. (Since the only universal structure available at $\mathcal{I}$ is this of a smooth manifold, we do not have any additional ingredient.)
    \item[(ii)] $\bf{\Theta}$ should be conformally invariant
    \item[(iii)] Since Einstein-Hilbert action depends analytically upon the metric, so should $\bf{\Theta}$.
\end{itemize}
It is easy to notice that \eqref{theta} satisfies them all. However, in no (possibly generalized) way property (iv) from the list of requirements from \cite{Wald:1999wa} holds. Indeed, $\eqref{theta}$ is not zero even when evaluated on the Schwarzschild-de Sitter solution. \\
${\bf\Theta}$ is defined up to an addition of an exact variation of a functional $W[g^{(0)}, T]$. This must be of the form:
\begin{equation}
    W[g^{(0)}, T] = \int_{\mathcal{I}} d^3 x \sqrt{g^{(0)}} w (g^{(0)}, T),
\end{equation}
where $w$ is a scalar with a conformal weight $-3$. Since it is supposed to be analytic, local and covariant, it admits an expansion into monomials in the following:
\begin{itemize}
    \item $g^{(0)}_{ab}$ ($+2$)
    \item $g^{(0)ab}$ ($-2$)
    \item $\sqrt{g^{(0)}} \epsilon_{abc}$ ($+3$)
    \item Ricci tensor\footnote{Notice that since $\mathcal{I}^+$ is three dimensional, the Ricci tensor encodes the whole curvature} of $g^{(0)}$ and its covariant derivatives ($0$)
    \item $T_{ab}$ and its covariant derivatives ($-1$),
\end{itemize}
where numbers in the brackets denote the weight of each component under a constant rescaling. The fact that we can restrict ourselves into such monomials will be explained in Appendix \ref{app_exp}. We will show that there is no not-trivial monomial with the rescaling weight $-3$ and so there is no analytic, conformal invariant functional. Note, that if one abandons analyticity, such functionals exist, with the simplest example being
\begin{equation}
    \int_{\mathcal{I}} d^3 x \sqrt{g^{(0)}} \sqrt{C_{abc}C^{abc}},
\end{equation}
where $C_{abc}$ is a Cotton tensor of $g^{(0)}$. \\
Let us start our argument by accounting for the factors  containing $\epsilon_{abc}$ in $w$. One has a simple formula for a product of $\epsilon_{abc}$ with itself
\begin{equation}
  \sqrt{g^{(0)}} \epsilon_{abc} \sqrt{g^{(0)}} \epsilon_{def} = \sum_\sigma (-1)^{|\sigma|} g_{a \sigma(d)} g_{b\sigma(e)} g_{c\sigma(f)},
\end{equation}
where $\sigma$ are all permutations of a three-element set. Thus, we can assume without loss of generality that there is at most one factor $\epsilon_{abc}$ in our expansion. \\
First, let us consider the case when there is at least one $T$. The most general form of $w$ is thus:
\begin{equation}
    w = D_{a_1} D_{a_2} \dots D_{a_k} T_{ab} P_{b_1 b_2 ... b_n} \textrm{\ and}  \left( \frac{n+k}{2} + 1 \right)\textrm{contractions}.
\end{equation}
Notice that this factorization does not need to be unique.
Let $\deg P$ be a weight of $P_{b_1 b_2 ... b_n}$. It satisfies 
\begin{equation}
    \deg P \le n,
\end{equation}
where the equality holds only when $P_{b_1 b_2 ... b_n}$ is built out of  $g_{ab}$ and $\sqrt{g}\epsilon_{abc}$ alone. On the other hand, we have
\begin{equation}
    -3 = -1 + \deg P - 2 \left( \frac{n+k}{2} + 1 \right) = -3 + \deg P - n - k \le -3 - k
\end{equation}
Thus, $k=0$, $\deg P = n$. Since we have $\frac{n+k}{2} + 1$ contractions, it implies that $n$ is an even number. Thus
\begin{equation}
P_{b_1 b_2\dots b_n} \sim g_{b_1 b_2} \dots g_{b_{n-1} b_n}
\end{equation}
(up to permutations of indices and a proportionality constant.)
But that means that $w$ is proportional to the trace of $T_{ab}$ which vanishes if the equations of motion are satisfied. \\
We are left with the case when considered functional is $T$-independent. Then, it would mean that $w$ is an analytical, scalar conformal invariant (of weight $-3$). 
We will again show that $w$ cannot be a monomial in fields which has a weight $(-3)$ under constant rescalings. Since it is supposed to have an odd weight, it must contain a factor $\sqrt{g}\epsilon_{abc}$. But then,  $\epsilon$ needs to be contracted with an object with three indices. Since both the metric and Ricci tensor have two indices, it means that $w$ must contain an odd number of covariant derivatives acting on the Ricci tensors. In particular, there is at least one $R_{de}$ on which acts an odd number of $D_{a_i}$. We thus can write
\begin{equation}
    w = \sqrt{g} \epsilon_{abc} D_{a_1}\dots D_{a_k} R_{de} P_{b_1\dots b_n} \textrm{\ and} \left(\frac{n+k+5}{2} \right) \textrm{contractions},
\end{equation}
where $k$ and $n$ are odd and even, respectively. Again $P$ satisfies
\begin{equation}
    \deg P \le n,
\end{equation}
when the equality holds when $P_{b_1\dots b_n}$ is built out of $g_{ab}$ alone. On the other hand, we have
\begin{equation}
    -3 = 3 + \deg P - (n+k+5) = -2 - k + \deg P - n \le -2 - k.
\end{equation}
It follows that $k=1$ and $n = \deg P$ (so it is built out of $g_{ab}$s alone) and $w$ reads
\begin{equation}
    w \sim \sqrt{g} \epsilon_{abc} D_{a_1}R_{de} \textrm{\ and\ } 3 \textrm{\ contractions}.
\end{equation}
As a result of those contractions, we could either have $w \sim \sqrt{g} \epsilon^{abc} D_a R_{bc}$ or $w \sim \sqrt{g}  \epsilon^{abc} D_a R g_{bc}$ and both possibilities vanish due to the symmetry of indices. \\
Thus, ${\bf \Theta}$ is unique at least on-shell\footnote{The difference between on-shell and off-shell here lies in the definition of $\mathcal{F}$. We could have restricted ourselves to traceless $T_{ab}$s from the very beginning and then uniqueness would hold also off-shell}. Since at the end of the day we are only interested in charges and fluxes only on $\overline{\mathcal{F}}$, there is no ambiguity. \\
Notice that in particular 'another' choice of $\bf{\Theta}$:
\begin{equation}
    {\bf\Theta}' = -\delta \left(T^{ab} \sqrt{g^{(0)}} \right) g_{ab} d^3 x = {\bf \Theta} - \delta \left( T^{ab}  g_{ab} \sqrt{g^{(0)}} d^3 x\right) = \bf{\Theta}
\end{equation}
is perfectly equivalent to the one we made.
\subsection{Special cases} \label{Special} \noindent
In this part, we will deal with spacetimes expressed  in the Bondi coordinates. Fortunately, there is a dictionary between Bondi and Fefferman-Graham gauges given in \cite{Compere:2019bua}. \\
In \cite{Kolanowski:2020wfg} flux of energy in the GR linearized around the de Sitter was calculated. To this end, we used a Killing vector $\partial_u$ which generates translation from one null-cone to another in the Bondi coordinates in which the formula for the metric tensor reads
\begin{equation}
    \mathring{g} = - \left(1 - \frac{\Lambda r^2}{3} \right) du^2 - 2dudr + r^2 \mathring{\gamma}_{AB} dx^A dx^B. \label{g}
\end{equation}
Boundary terms were fixed by imposing invariance with respect to certain asymptotic symmetries dubbed superpseudotranslations. It yields
\begin{equation}
    \mathbf{F}_{\partial_u} = \frac{1}{16\pi G \ell} \mathcal{E}_{cd} \mathcal{L}_{\partial_u} g^{(0)}_{ab} \mathring{g}^{(0)ac} \mathring{g}^{(0)bd} \sqrt{\mathring{g}^{(0)}} d^3 x, \label{fluxds}
\end{equation}
where $\mathring{g}^{(0)}$ is a metric induced on $\mathcal{I}^+$ without any perturbation. Looking at \eqref{weyl}, it is evident that this expression is Taylor expansion of \eqref{flux} up to the quadratic terms. (The difference of signs is compensated by orientation of $\mathcal{I}$) It was also shown in \cite{Kolanowski:2020wfg} that \eqref{fluxds} reduces to the Trautman-Bondi mass loss law in the limit $\Lambda \to 0$ which suggests that this is the correct formula for the energy {\it density} at null infinity. \\
Let us now look at the Schwarzchild-de Sitter (also known as Kottler) metric:
\begin{equation}
    \mathring{g} = - \left(1 - \frac{\Lambda r^2}{3} - \frac{2M}{r} \right) du^2 - 2dudr + r^2 \mathring{\gamma}_{AB} dx^A dx^B.
\end{equation}
Initial data on $\mathcal{I}$ reads \cite{Ashtekar:2014zfa}
\begin{align}
    g^{(0)} &= du^2 + \ell^2 \mathring{\gamma}_{AB} dx^A dx^B, \\
    \mathcal{E} &= -3GM \ell^{-3} \left(du^2 - \frac{1}{3}g^{(0)} \right).
\end{align}
where $\mathring{\gamma}$ is a standard metric on a unit sphere. This metric $g$ is equipped with a $4$-dimensional space of the Killing vectors. This algebra contains a one-dimensional center generated by $\partial_u$ which corresponds to the 'time'-translation (and thus 'energy') and unique three-dimensional subalgebra $so(3)$ of rotations generating angular momenta. Notice that those come with a natural scaling given by a commutator:
\begin{equation}
    [R_i, R_j] = \epsilon_{ijk} R_k.
\end{equation}
The scaling of the time-translation generator $\partial_u$ can be fixed because metric $g^{(0)}$ coincides with $g^{(0)}$ of the de Sitter spacetime. Note that the fact that $\mathcal{E} \neq 0$ on this background changes a little bit the nature of fluxes. If we consider a purely magnetical gravitational wave (it means, a perturbation which changes a conformal class of $g^{(0)}$ but not $\mathcal{E}$), we would obtain non-zero $\mathcal{F}$. Nevertheless, this addition integrates to zero over the whole $\mathcal{I}^+$ and vanishes in the limit $\Lambda \to 0$. \\
Now we will  comment on how to define energy on the Kerr-de Sitter (also known as Carter) background.  Kerr-de Sitter metric is equipped with a $2$-dimensional abelian algebra of Killing vectors generated by $\partial_u, \partial_\phi$. We want to find appropriate generators. 1-dim. algebra generated by $\partial_\phi$ possesses a unique property that its integral curves on $\mathcal{I}$ are closed\footnote{A priori one could also close integral curves of $\partial_u$ by saying that $u$ is periodic. This would obviously change topology of the whole spacetime. Nevertheless, even then $\partial_\phi$ is distinguished because it vanishes at some points.} and thus could be rotation generator. We can fix its 'length' by demanding that period of $\phi$ is $2\pi$. There is nothing distinguished about $\partial_u$. The only geometric condition to be satisfied on $\mathcal{I}$ we were able to think of to somehow choose 'time'-translation generator $T$ was
\begin{equation}
    g^{(0)} (T, \partial_\phi) = 0
\end{equation}
since it is a conformally invariant requirement. It turns out, that the conformal metric defined on $\mathcal{I}$ admits that possibility. We discuss it in detail now. According to \cite{Ashtekar:2014zfa}, the conformal metric tensor induced on $\mathcal{I}$ is
\begin{equation} g^{(0)}=\frac{du^2 - 2a\rm sin^2\theta du d\phi}{(1+\frac{a^2}{\ell^2})^2}  + \frac{\ell^2\rm sin^2\theta d\phi^2}{1+\frac{a^2}{\ell^2}} + 
\frac{\ell^2d\theta^2}{1+\frac{a^2}{\ell^2\rm cos^2\theta}} \end{equation}
Then, (up to a rescaling)
\begin{equation}
    T = \partial_u + \frac{a}{a^2 + \ell^2}\partial_\phi.
\end{equation}
Such $T$ has the correct limit as $\Lambda \to 0$. To learn more about the vector field $T$ let us  transform the metric tensor to the coordinate system $(u,\theta,\phi')$ adapted to $T$, namely such that
\begin{equation}
T^a\partial_a\phi' = 0, \ \ \ \ \phi' = \phi - \frac{a}{\ell^2 + a^2}u.
\end{equation}
Then
\begin{equation} 
g^{(0)}=\frac{1 + \frac{a^2}{\ell^2}\cos^2 \theta}{(1+\frac{a^2}{\ell^2})^3}du^2
 + \frac{\ell^2\rm sin^2\theta }{1+\frac{a^2}{\ell^2}}d\phi'^2 + 
\frac{\ell^2}{1+\frac{a^2}{\ell^2\rm cos^2\theta}}d\theta^2,\end{equation}
while
\begin{equation} 
T = \partial_u.
\end{equation}
Now we can see, that the Killing vector $T$ is non-twisting - it is orthogonal to the surfaces 
\begin{equation}
u = \rm const .     
\end{equation}
Nevertheless, it is unclear to us whether some physical reasons are standing behind this particular choice. Moreover, noticed that since $g^{(0)}_{Kerr} \neq g^{(0)}_{dS}$, we do not have a tool to fix the scaling of the energy. \\
Finally, let us consider more general background given by $[g^{(0)}, \mathcal{E}]$. One could ask when, given a representative $(g^{(0)}, \mathcal{E})$ this structure (and thus the whole spacetime) possess certain symmetry. In other words, when there are a positive function $\Omega: \mathcal{I} \to \mathbb{R}_{>0}$ and a vector field $K$ such that
\begin{align}
    \begin{split}
        \mathcal{L}_K \Omega^2 g^{(0)} &= 0 \\
        \mathcal{L}_K \Omega^{-1} \mathcal{E} &= 0.
    \end{split}
\end{align}
The answer is relatively simple. It follows that $K$ must be a conformal symmetry of both $(g^{(0)}, \mathcal{E})$:
\begin{align}
    \begin{split}
                \mathcal{L}_K g^{(0)} &= \alpha g^{(0)}  \\
        \mathcal{L}_K \mathcal{E} &= \beta \mathcal{E},
    \end{split}
\end{align}
where $\alpha = -2\beta$. Since our Hamiltonians and fluxes are conformally invariant, we do not need to find $\Omega$ -- knowledge of its existence is enough.
\section{Summary and discussions} \label{con} \noindent
In this work, we have introduced 'conserved' charges associated with the diffeomorphism of null infinity in the asymptotically de Sitter spacetimes. We have shown that our definition is unambiguous under certain natural conditions. Moreover, we have checked that in the perturbative regime the limit $\Lambda\to0$ reproduces familiar expressions for the Trautman-Bondi mass. Nevertheless, a few questions remain. First of all, what is the physical meaning of the charges and fluxes associated with a diffeomorphism of $\mathcal{I}^+$ (if any)? For now, we were only able to answer that question for the isometries of certain backgrounds. The general answer is still lacking. Even in the case of the Carter spacetime, we do not find any physical (in contrast to geometrical) guide to distinguish the 'time'-translation generator. The harder question of whether one can do it at all in the full theory is currently being investigated in detail. Since $\Lambda$ is extremely small, it is not the question one needs to answer to extract even $\Lambda$-dependent effects from the observations of the gravitational waves but rather one of a more fundamental character.
\\
From the point of view of a quantum theory, it would be beneficial to have a Poisson bracket $\lbrace \cdot, \cdot \rbrace$ between just introduced observables. However, this would require inverting symplectic form. Since we are in the infinite-dimensional phase space, it is rather a delicate task. One should understand the topology of $\mathcal{F}$ first. For example, phase space in \cite{Ashtekar:1981bq} was taken to be a manifold modeled on a Frechet space. In our case situation seems to be even more complicated. On the other hand, it seems to be easier than in the usual initial value problem so we hope it should be manageable. Indeed, taking care of an algebraic constraint 
\begin{equation}
    g_{ab}^{(0)} T^{ab} = 0
\end{equation}
is rather automatic and so one is left only with a condition
\begin{equation}
    D_a T^{ab} = 0.
\end{equation}
We hope to adress it in the future. \\
Let us now compare our results with the ones presented so far in the literature. As we have mentioned, Eq. \eqref{hamiltonian} was obtained for the first time in \cite{Anninos:2010zf} but an argument for the uniqueness was lacking. We have filled this gap in Sec. \ref{uniqueness}. Analogous results were also derived in \cite{Ashtekar:2014zfa, Balakrishnan:2019zxm} but under much stricter boundary conditions. Namely, it was assumed that $g_{ab}^{(0)}$ was conformally flat and that $\xi$ is a conformal Killing vector field of $[g^{(0)}]$. Then, it was found that all 10 fluxes vanish in the agreement with Eq. \eqref{flux}. Finally, in \cite{Compere:2020lrt} it was assumed that $\mathcal{I}^+$ is equipped with a distinguished foliation. This restricts the group of the asymptotic symmetries from all diffeomorphisms of $\mathcal{I}^+$ to only those which preserve this foliation. Associated charges were calculated in Sec. 2 of \cite{Compere:2020lrt} and they agree with our results (restricted to the smaller algebra). A presented argument for their uniqueness was entirely different than the one from this work. It was shown that it is the only possible choice of Hamiltonians (given by a split into 'integrable' and 'non-integrable' parts of \eqref{cancel}) which makes the charge algebra closed under the adjusted Dirac bracket \cite{Barnich:2011mi}. This agreement is quite remarkable, given that there is still no derivation of that bracket from the covariant symplectic formalism in this context. However, in Sec. 4 authors noticed that this expression (in a radial Bondi gauge) has an ill-defined limit as $\Lambda \to 0$ which led them to the change of considered boundary terms in Lagrangian\footnote{It is by no means in contradiction with the results of Sec. \ref{uniqueness}  because that addition depends explicitly upon the distinguished foliation.} and thus different presymplectic form, potential and so also Hamiltonians and fluxes. Divergences in the limit $\Lambda \to 0$ are in tension with \cite{Chrusciel:2002cp, Kolanowski:2020wfg} where this limit was explicitly checked (for the linearized gravity) but in an 'anti-radial' Bondi gauge. Thus it seems that this limit may depend non-trivially on the choice of gauge\footnote{Since we consider transformations at $\mathcal{I}^+$, there is an additional subtlety, whether solutions considered in \cite{Compere:2020lrt} and in \cite{Chrusciel:2020rlz, Kolanowski:2020wfg} differ only by a gauge or rather an asymptotic symmetry} and further investigations are needed.
\\
To conclude, we have (re-)discovered charges and fluxes in the asymptotically de Sitter spacetimes. Without any doubt, those can be used while working within the linearized gravity (on a suitably symmetrical background) to define physical quantities. As we remarked in the Introduction, thanks to the relative simplicity of the initial value problem at $\mathcal{I}^+$ \cite{friedrich1986}, it is also possible to construct a canonical phase space at $\mathcal{I}^+$ by imposing suitable fall-off conditions on the initial data as one approaches $i^+$ and $i^o$ \cite{aa-sanya}. As in the standard ADM phase space description in asymptotically flat space-times, the boundary conditions make it possible to single out symmetries from gauge, and define 2-sphere `charges' at $i^+$ and $i^o$ and global fluxes across $\mathcal{I}^+$ associated with the symmetries. This suggests that it should be possible to combine advantages of the `local' approach adopted here with those of the `global' approach adopted in \cite{aa-sanya} to obtain also `local'  2-sphere charges and fluxes associated with symmetries. This issue is under investigation. Having succeed in understanding this relationship, one could try to quantize the system using proposed 'conserved' quantities as building blocks for the observables-to-be. We hope to address it in the future.
\acknowledgments
We thank Abhay Ashtekar, Piotr Chruściel and Tomasz Smołka for fruitful discussions. MK was financed from budgetary funds for science in 2018-2022 as a research project under the program "Diamentowy Grant". JL was supported by Project OPUS 2017/27/B/ST2/02806 of  Polish National Science Centre.
\appendix
\section{From analytical functionals to monomials} \label{app_exp}
Consider the tensors  
\begin{equation}g^{(0)}_{ab},\ T_{ab},\ R_{ab},\end{equation} 
and their covariant derivatives 
\begin{equation}D_{a_1}...D_{a_k}T_{bc},\ \ \ D_{a_1}...D_{a_l}R_{bc}\end{equation} 
to an arbitrary order at a point  
\begin{equation}x\in \mathcal{I}.\end{equation} 
Let 
\begin{equation}f(g^{(0)}_{ab},T_{ab},R_{ab},D_{a_1}...D_{a_k}T_{bc},D_{a_1}...D_{a_l}R_{bc})\end{equation} 
be an invariant. Suppose that $f$ can be expanded to perhaps an infinite series of finite monomials
\begin{equation}f = ... +   P_{a_1...a_n}{}^{b_1...b_m} + ... \end{equation}
in the arguments of $f$\footnote{For example $R=... +g^{(0)}{}^{12}R_{12}+...$}. Finally, suppose the expansion commutes with the Haar integral of $f$ over the compact group  $SO(g^{(0)})$ of the rotations of the tangent space $T_x{\mathcal{I}}$ endowed with the metric $g^{(0)}$. Clearly, the invariant $f$ is insensitive on the integration with respect to the rotations (the measure is normalized to $1$). On the other hand, we may perform the integral term by term. For each of them, the result takes the following form
\begin{align} &\int d\mu_H(h)P_{a'_1...a'_n}{}^{b'_1...b'_m} h^{-1}{}^{a'_1}_{a_1}...h^{-1}{}^{a'_n}_{a_n}
h_{b'_1}^{b_1}...h_{b'_m}^{b_m}\nonumber\\ 
&= P_{a'_1...a'_n}{}^{b'_1...b'_m} \sum_l \iota_{(l)}{}^{a'_1...a'_n}{}_{b'_1...b'_m} \iota_{(l)}{}_{a_1...a_n}{}^{b_1...b_m}  \end{align}
where each factor 
\begin{equation}\iota_{(l)}{}^{a'_1...a'_n}{}_{b'_1...b'_m}\end{equation}
is an invariant tensor.  Consider a term 
\begin{equation} P_{a'_1...a'_n}{}^{b'_1...b'_m}  \iota_{(l)}{}^{a'_1...a'_n}{}_{b'_1...b'_m}  \end{equation}
The invariant $\iota_{(l)}$ is a sum of products of the elementary $SO(g^{(0)})$ invariants: 
\begin{equation}g_{a'_ia'_j},\ g_{a'_i}{}^{b'_j},\ g^{b'_ib'_j},\ \sqrt{g^{(0)}}\epsilon_{a'_ia'_ja'_k},\ \sqrt{g^{(0)}}\epsilon_{a'_ia'_j}{}^{b'_k},\ \sqrt{g^{(0)}}\epsilon_{a'_i}{}^{b'_jb'_k},\ \sqrt{g^{(0)}}\epsilon^{b'_ib'_jb'_k}.\end{equation} 
The  $\epsilon$ terms can be absorbed into $P$. Then, the result takes the form of a monomial with all the indices contracted with $g_{a'_ia'_j},\ g_{a'_i}{}^{b'_j}$, and/or $g^{b'_ib'_j}$ considered in Section \ref{uniqueness} above. 
 \bibliographystyle{plain}
\bibliography{bibl.bib}
\end{document}